\documentclass{article}

\usepackage{amsmath}
\usepackage{latexsym}
\usepackage{amssymb}
\usepackage{bm}
\usepackage{graphics,epstopdf}
\usepackage{color}
\usepackage{hyperref}

\usepackage{newlfont}
\usepackage{amsfonts}
\usepackage{amsthm}
\usepackage{graphicx}
\usepackage{epsfig}

\newcommand{\ket}[1]{|{#1}\rangle}
\newcommand{\bra}[1]{\langle{#1}|}
\newcommand{\braket}[2]{\langle {#1} | {#2} \rangle}
\newcommand{\ketbra}[2]{|{#1} \rangle \langle {#2} |}
\renewcommand{\t}[1]{\textrm{#1}}
\newcommand{\com}[1]{[\textsc{#1}]}
\usepackage{times}

\newcommand*{\Perm}[2]{{}^{#1}\!P_{#2}}%
\newcommand*{\Comb}[2]{{}^{#1}C_{#2}}%
\usepackage[up]{subfigure}

\newcommand{\be}{\begin{equation}}
\newcommand{\ee}{\end{equation}}
\newcommand{\bc}{\begin{center}}
\newcommand{\ec}{\end{center}}
\newcommand{\bea}{\begin{eqnarray}}
\newcommand{\eea}{\end{eqnarray}}
\newcommand{\ba}{\begin{array}}
\newcommand{\ea}{\end{array}}
\newcommand{\tr}{\textcolor{red}}
\newcommand{\tb}{\textcolor{blue}}

\begin{document}

\title{\bf{Conditions for Anomalous Weak Value}}

\author{Arun Kumar Pati$^{(1)(2)}$ and Junde Wu$^{(1)}$\\
\small{$^{(1)}$Department of Mathematics}\\
\small{Zhejiang University, Hangzhou 310027, PR~ China}\\
\small{and}\\ 
\small{$^{(2)}$Quantum Information and Computation Group,}\\
\small{Harish-Chandra Research Institute,}\\
\small{Allahabad 211 019, India}}

\maketitle

\begin{abstract}
We show that the weak value of any observable in pre- and post-selected states can be expressed as the sum of the 
average of the observable in the pre-selected state and an anomalous part. We argue that at a fundamental level the 
anomalous nature of the weak values arises due to the interference between the
post-selected state and another quantum state which is orthogonal to the pre-selected state. This provides a
necessary and sufficient condition for the anomalous nature of the weak value of a quantum 
observable. Furthermore, we prove that for two non-commuting observables the product of their anomalous parts cannot be arbitrarily 
large.
\end{abstract}



\vskip 1cm

The concept of weak measurement and weak value have played a fundamental role in quantum theory since their inception.
The idea was first introduced by Aharonov-Albert-Vaidman
\cite{aha} while investigating the properties of a quantum system in  pre and post-selected ensembles.
 If the system is weakly coupled to an apparatus, then upon post-selection of the system onto a final state, the apparatus variable is shifted by 
 the weak value.
Unlike the usual quantum mechanical average of an observable, the weak value of an observable can have strange properties \cite{duck}.
In general, the weak value can be a complex number and can be large.
Moreover, it can take values outside the spectrum of the observable being measured. 
The concept of weak value is no more a theoretical prediction 
but has found numerous practical applications  in recent years \cite{nw,gj,oh,dixon,hof}. 
This provides us a new tool to look at complementary aspects of quantum quantum world \cite{av,sandu1,sandu2,wu}.

Admittedly, the weak value of any observable can be anomalous, in the sense that it can take value beyond the expectation value
the observable (either in the pre-selected or post-selected states). However, there is no physical explanation of the origin of the
anomalous nature of the weak value and how precisely the weak value differs from the average value of the observable. In this note, we give a simple
explanation of the origin of the anomalous nature of the weak value.
First, we show that the weak value for an observable in any pre and post-selected states can be expressed as the sum of the average of the observable in the pre-selected state and an anomalous part. This explains how at a fundamental level, the anomalous weak value arises  due to the interference of amplitudes for the post-selected state and another quantum state which is orthogonal to the pre-selected state. This gives us a necessary and sufficient condition for the weak value to be anomalous in nature.
We provide a lower bound on the extent to which the weak value
differs from the average value of the observable. Interestingly, we prove that for two non-commuting observables the product of 
the modulus of the anomalous 
part of their weak values cannot be arbitrarily large. This implies a new limitation on the preparation of pre- and post-selected 
quantum states. We also provide an upper bound on the difference between the real part of the weak value and
the largest eigenvalue of the observable.

{
In the weak measurement formalism, we start a quantum system which is preselected in the state $\ket{\psi_i} = \ket{\psi} \in {\cal H}_S$ and an apparatus in the state $\ket{\Phi} \in {\cal H}_A$.
The weak measurement can be realized using the interaction between the system and the
measurement apparatus which is governed by the interaction Hamiltonian
\begin{align}
 H_{int}= f(t) A \otimes M,
\end{align}
where $f(t)$ is the strength of the interaction with $\int f(t)dt= g$, $A$ is an observable of the system and $M$ is that of the apparatus (often called meter variable).
This is the von Neumann model of measurement when the coupling strength is arbitrary. But if $g$ is small, then we can realize the weak
measurement of an observable $A$ of the system. The interaction Hamiltonian allows the initial state of the system and apparatus $\ket{\psi} \otimes \ket{\Phi}$ to evolve as
\begin{align}
 \ket{\psi} \otimes \ket{\Phi}  \rightarrow e^{-i g A \otimes M} \ket{\psi} \otimes \ket{\Phi}.
\end{align}
After the weak interaction, we post-select the system in the state $\ket{\phi}$ with the
postselection
probability given by $p= |\braket{\phi}{\psi}|^2 (1 + 2 g Im \langle A\rangle_w \langle M \rangle)$, where
$\langle M \rangle = \braket{\Phi}{M |\Phi }$ and the weak value of $A$ is
given by
\begin{align}
\langle A\rangle_w = \frac{\braket{\phi}{A |\psi }}{\braket{\phi}{\psi} }.
\end{align}
The effect of the weak interaction and post-selection is that at intermediate time between two strong measurements the apparatus 
state is subject to an effective Hamiltonian $H = g \langle A\rangle_w M$ and the initial state of apparatus changes as
\begin{align}
\ket{\Phi}  \rightarrow e^{-ig \langle A\rangle_w  M} \ket{\Phi}.
\end{align}
Now, using the Vaidman formula \cite{vaid} we can express the action of the Hermitian operator $A$ on the pre-selected state  
 as
\begin{align}
A \ket{\psi }  = \langle A  \rangle \ket{\psi} +  \Delta A \ket{\bar{\psi}},
\end{align}
where $\langle A \rangle =  \bra{\psi} A   \ket{\psi }$,
$\Delta A$ is the uncertainty in the state $\ket{\psi}$, i.e.,
$\Delta A^2 = \langle \psi|(A-\langle A\rangle)^2 |\psi \rangle$ and $\ket{\bar{\psi}} \in {\bar {\cal H}}$ is a state
orthogonal to $\ket{\psi}$.
With this the weak value of an observable $A$ can be expressed as
\begin{align}
\label{vaid}
\langle A\rangle_w = \langle A \rangle  +  \Delta A \frac{ \braket{\phi}{ {\bar{\psi}}} }{ \braket{\phi}{\psi} } .
\end{align}
This shows that the weak value in any pre- and post-selected states can be expressed as the sum of the average of the observable 
in the pre-selected state and an anomalous part, where the anomalous part is defined as 
\begin{align}
\delta \langle A\rangle_w =  \Delta A \frac{ \braket{\phi}{ {\bar{\psi}}} }{ \braket{\phi}{\psi} } .
\end{align}

The above formula is very important in understanding the anomalous nature of the weak value. Note that by the formalism of the weak
measurement, $ \braket{\phi}{\psi} \not= 0$, though it can be very small in principle (small but finite). Otherwise, the success
probability of the post-selection is zero, and we have nothing to talk about. By choosing small value of the overlap between the pre- and post-selected states one can make the weak value large, but this gives only a sufficient condition for the anomalous nature of the 
weak value. Even if $ \braket{\phi}{\psi}$ is small, if $\braket{\phi}{ {\bar{\psi}}} = 0$, we cannot have a weak value beyond the 
 average of the observable. Therefore, to have a weak value that goes
beyond the average of the observable, we must have nonzero uncertainty and non-zero value of $\braket{\phi}{ {\bar{\psi}}}$. Also, it is
not enough to have an uncertainty in $A$ in the pre-selected state, but we should have the ability to produce an 
interference between $\ket{\phi}$ and
$ \ket{\bar{\psi}}$. In quantum theory, two orthogonal states cannot interfere, and hence if $\braket{\phi}{ {\bar{\psi}}}$ is non-zero
then they have the ability to interfere.  Thus, two non-orthogonal states can interfere
and when they are in phase, that will yield a maximum interference pattern. The `in phase' criterion is due to Pancharatnam \cite{sp} which
says that any two non-orthogonal vectors $\ket{\psi_1}$ and $\ket{\psi_2}$ are in phase if $\braket{\psi_1}{ \psi_2} $ is
real and positive (non-zero). Therefore, when $\braket{\phi}{ {\bar{\psi}}} = 0$, then the weak value will be equal to the average
 of the observable in the pre-selected state and it will be in the range of the eigenvalues
of the observable $A$. Hence, the necessary and sufficient condition for the anomalous weak value of an observable
 is that the post-selected state and the state
orthogonal to the pre-selected state should not be orthogonal and should interfere. It is the ability of these two states to interfere  that leads to a weak value beyond the average value of the observable.

The average of an observable $A$ in a quantum state $\ket{\psi}$ can be expressed as an average of intermediate weak values. Note that by inserting
a resolution of identity $\sum_k \ket{\psi_k}{\bra {\psi_k} } = I$ we can write
the average of $A$ as 
\begin{align}
\langle \psi| A| \psi \rangle =  \sum_k  |\braket{\psi}{\psi_k}|^2  \langle A \rangle_w^{(k)},
\end{align}
where $\langle A \rangle_w^{(k)} = \frac{\braket{\psi_k}{A |\psi }}{\braket{\psi_k}{\psi} }$ is the weak value corresponding to $k$th 
post-selection. It is interesting to see that even though each of these $\langle A \rangle_w^{(k)}$ is a complex number, still the average gives a real number. To understand the underlying mechanism, note that for each $k$, the weak value can be
written as a sum of the average of $A$ in the state $\ket{\psi}$ and an anomalous part $\delta \langle A \rangle_w^{(k)} = 
\Delta A \frac{ \braket{\psi_k}{ {\bar{\psi}}} }{ \braket{\psi_k}{\psi} } $. Therefore, all these intermediate weak values do remain outside 
the range of the spectrum of the observable $A$. But then why we do not see their contributions? The answer is that the average interference turns 
out to be exactly equal to the overlap between two orthogonal states and we do not see the contributions of the anomalous parts of the intermediate weak values, i.e., precisely we have $\sum_k |\braket{\psi}{\psi_k}|^2  \delta \langle A \rangle_w^{(k)} = 0$.

Second, we can understand more deeply why the weak value is complex. Note that complex nature of the weak value comes solely from the second term.
To see this clearly, we can express the real and imaginary parts of the weak value as
\begin{align}
Re \langle A\rangle_w &= \langle A \rangle  +  \Delta A \frac{ |\braket{\phi}{ {\bar{\psi}}}| }{ |\braket{\phi}{\psi}| } \cos(\bar{\Phi} - \Phi) \nonumber\\
Im\langle A\rangle_w &=  \Delta A \frac{ |\braket{\phi}{ {\bar{\psi}}}| }{ |\braket{\phi}{\psi}| } \sin(\bar{\Phi} - \Phi),
\end{align}
where $\bar{\Phi}$ is the relative phase difference between $\ket{\phi}$ and $ \ket{\bar{\psi}}$, and   $ \Phi$ is the relative phase difference
between $\ket{\phi}$ and $ \ket{\psi}$. This shows that not only it is the interference between the pre and post-selected states, but
most importantly, it is the interference between the post-selected state and the orthogonal state to the pre-selected state which plays a
crucial role in deciding the anomalous behaviour of the weak value. If $(\bar{\Phi} - \Phi) = n \pi$, with $n$ being an integer, then
$Im \langle A\rangle_w =0$. This means, the weak value can be anomalous without being complex. If $(\bar{\Phi} - \Phi)$ is an odd multiple of $\pi$, then $Re \langle A\rangle_w < \lambda_{min}(A)$, where $\lambda_{min}(A)$ is lowest eigenvalue of
$A$. This in turn implies that the real part of the weak value will be outside the spectrum of the observable. If $(\bar{\Phi} - \Phi)$ is an odd multiple of $\frac{\pi}{2}$, then $Re \langle A\rangle_w = \langle A\rangle$ and $Im\langle A\rangle_w =  \Delta A \frac{ |\braket{\phi}{ {\bar{\psi}}}| }{ |\braket{\phi}{\psi}| }$ and it can still be anomalous.

Now, we will show how much the weak value differs from the average of the corresponding observable. Using (\ref{vaid}) and the Schwarz inequality,
we have
\begin{align}
|\langle A\rangle_w  - \langle A \rangle| \ge     \Delta A | \braket{\phi}{ {\bar{\psi}}}|.
\end{align}
Thus, the modulus of the anomalous part of the weak value $|\delta \langle A\rangle_w| = |\langle A\rangle_w  - \langle A \rangle|$ is bounded below by the quantum uncertainty in the observable and the overlap of the post-selected state and another state orthogonal to
the pre-selected state. In fact, the anomalous part of the weak value satisfies the following inequality 
\begin{align}
\Delta A | \braket{\phi}{ {\bar{\psi}}}| \le |\delta \langle A\rangle_w |  \le \frac{ \Delta A }{ |\braket{\phi}{\psi}| }.
\end{align}
Interestingly, if we have two non-commuting observables $A$ and $B$, then one can have a tradeoff relation for the
anomalous part of their weak values. Using the Robsertson uncertainty relation for two incompatible observables \cite{rob} we have
\begin{align}
|\delta \langle A\rangle_w|~~ |\delta \langle B\rangle_w|  \ge     \frac{1}{2} | \braket{\psi}{[A, B] |\psi} | 
| \braket{\phi}{ {\bar{\psi}}_A }|  | \braket{\phi}{ {\bar{\psi}}_B }|.
\end{align}
For two canonical conjugate observables like position and momentum of the system the product of their respective anomalous
parts will satisfy
\begin{align}
|\delta \langle X\rangle_w|~~ |\delta \langle P\rangle_w|  \ge     \frac{\hbar}{2} 
| \braket{\phi}{ {\bar{\psi}}_X }|  | \braket{\phi}{ {\bar{\psi}}_P }|.
\end{align}
This shows that for two canonical pair of observables, for the same pre- and post-selected ensemble, the anomalous part of their weak
values cannot be arbitrarily large. If the anomalous part of the weak value for the position observable is large, then correspondingly, the the anomalous part of the weak value for the momentum will be small. This imposes a new limitation on the preparation of pre- and post-selected quantum states. This can have  practical applications. For example, if we aim to amplify small signals using weak values corresponding to two
complementary observables then there will be a limitation. We cannot amplify both the signals at the same time.

We can also provide an upper bound on the difference between the real part of the weak value and the largest eigenvalue of the observable. First, note that the overlap between the pre and post-selected state can be made real. Given any pre and post-selected (PPS) ensemble
$\{ \bra{\phi}, \ket{\psi} \}$, we can always define another PPS ensemble $\{ \bra{\phi}, \ket{\chi} \}$, where
$\ket{\chi} = \frac{ \braket{\psi}{\phi} }{ |\braket{\psi}{\phi}| } \ket{\psi}$ with $\braket{\chi}{\phi}$ as real. Since
$\ket{\psi}$ and $\ket{\chi}$ differs by an overall complex number of unit modulus, they are equivalent pre-selected quantum states, i.e., 
$\ket{\psi} \sim \ket{\chi}$.
Hence, the weak value $\langle A\rangle_w = \frac{\braket{\phi}{A |\psi }}{\braket{\phi}{\psi} }=  \frac{\braket{\phi}{A |\chi }}{\braket{\phi}{\chi} }$.
With this equivalent PPS, we can express the weak value as
\begin{align}
\label{vaid3}
\langle A\rangle_w = \langle A \rangle  +  \Delta A \frac{ \braket{\phi}{ {\bar{\chi}}} }{ |\braket{\phi}{\psi}| },
\end{align}
where $\ket{\bar{\chi}}$ is orthogonal to $\ket{\chi}$. From the above expression it is clear that the complex nature of 
the weak value arises solely from $\braket{\phi}{ {\bar{\chi}}} $.  Therefore, if $\ket{\phi}$ and $\ket{ {\bar{\chi}}} $ are in phase then they will produce maximal interference.
In that case, we will have $\langle A\rangle_w$ real (though still it can be anomalous).
Using the expression (\ref{vaid3}) and the fact that $Re \braket{\phi}{ {\bar{\chi}}} \le |\braket{\phi}{ {\bar{\chi}} }|$, we have
\begin{align}
Re \langle A\rangle_w  -\lambda_{max}(A) \le   \frac{\Delta A} { |\braket{\phi}{\psi} |},
\end{align}
where $\lambda_{max}(A)$ is the largest eigenvalue of the observable.

To conclude, we have shown that the weak value for an observable in any pre and post-selected states is the sum of the average of the observable in the pre-selected state and an anomalous part. We have argued that the anomalous weak value arises due to the interference of the amplitudes for the post-selected state and another states which is orthogonal to the pre-selected state. This provides a necessary and sufficient condition for the anomalous nature of the weak value.
We have provided a lower bound for the difference between the weak value
and the average value of the observable. This results in a new limitation on the anomalous part of the weak value.
We have proved that for two non-commuting pair of observables, for the same pre- and post-selected ensemble, the anomalous part of their weak
values cannot be arbitrarily large. 
Our result also sheds light on why in the classical world we cannot have the anomalous weak value.
In the classical world, unless we deal with waves directly, we do not have the notion of interference of 
different amplitudes. Thus, any classical model cannot have the ability to mimic the interference between the post-selected state and  another state orthogonal to the pre-selected state, and hence no place for anomalous weak values.

\subsubsection*{Acknowledgement}

This work is supported by National Natural Science
Foundation of China (11171301 and 10771191) and the Doctoral Programs
Foundation of Ministry of Education of China (J20130061), and it is also supported
by the Special Project of University of Ministry of Education of China and the
Project of K. P. Chair Professor of Zhejiang University of China.


\begin{thebibliography}{999}



\bibitem{aha} Y. Aharonov, D. Z. Albert, and L. Vaidman, Phys. Rev. Lett. {\bf 60}, 1351 (1988). 	





\bibitem{duck} I. M. Duck, P. M. Stevenson, and E. C. G. Sudarshan, Phys. Rev. D {\bf 40} 2112 (1989).

\bibitem{nw}  N. W. M. Ritchie, J. G. Story, and R. G. Hulet, Phys. Rev. 	
Lett. {\bf 66}, 1107 (1991). 

\bibitem{gj}  G. J. Pryde, J. L. O'Brien, A. G. White, T. C. Ralph, and 	
H. M. Wiseman, Phys. Rev. Lett. {\bf 94}, 220405 (2005).

\bibitem{oh}  O. Hosten and P. Kwiat, Science {\bf 319}, 787 (2008). 	

\bibitem{dixon}  P. B. Dixon, D. J. Starling, A. N. Jordan, and J. C. Howell, 	
Phys. Rev. Lett. {\bf 102}, 173601 (2009). 	


\bibitem{hof} H. H. Hofmann, Phys. Rev. A {\bf 83}, 022106 (2011).


\bibitem{av} Y. Aharonov and L. Vaidman, Lect. Notes Phys. {\bf734}, 399 (2008).

\bibitem{sandu1} S. Popescu, Physics {\bf 2}, 32 (2009).

\bibitem{sandu2} Y. Aharonov, S. Popescu, and J. Tollaksen, Phys. Today {\bf 63}, 27 (2010).

\bibitem{wu} S. Wu, Sci. Rep. {\bf 3}, 1193 (2013).

\bibitem{vaid} L. Vaidman, 
Am. J. Phys. {\bf 60}, 182 (1992).

\bibitem{sp} S. Pancharatnam 
Proc. Indian Acad. Sci. A {\bf 44} 247 (1956).

\bibitem{rob} H. P. Robertson, 
Phys.  Rev. {\bf 34}, 163 (1929).

\end{thebibliography}
\end{document}